\documentstyle[seceq,mbf,epsf]{ptptex}

\def\dsp{\displaystyle}
\def\mymatrix#1#2#3#4{\left( \matrix{\dsp#1&\dsp#2 \cr
                                     \dsp#3&\dsp#4 \cr} \right) }
\def\sanmatrix#1#2#3#4#5#6#7#8#9{\left( \matrix{\dsp#1&\dsp#2&\dsp#3 \cr
               \dsp#4&\dsp#5&\dsp#6 \cr \dsp#7&\dsp#8&\dsp#9 \cr} \right) }
\def\myvector#1#2{\pmatrix{\dsp#1 \cr \dsp#2  \cr}}

\def\cspan{\rule[-2.2ex]{0pt}{5ex}}
\def\pl{P}
\def\R{{\rm R}}

\def\m1{{-1}}
\def\ncdot{\hspace{-2pt}\cdot\hspace{-3pt}}
\def\ff{f}
\def\QQ{{\hspace{1pt}x}}
\def\VT{{V'}}
\def\VEV#1{\langle#1\rangle}
\newcommand{\nn}{\nonumber\\}

\preprintnumber{
KUNS-1557\\ hep-ph/9902204}

\title{Neutrino Masses in ${\mbf E_6}$ Unification}

\author{
Masako {\sc Bando}\footnote{E-mail: bando@aichi-u.ac.jp}
and Taichiro {\sc Kugo}$^{1,}$\footnote{E-mail:
kugo@gauge.scphys.kyoto-u.ac.jp}
}

\inst{%
Physics Division, Aichi University, Aichi 470-0296, Japan
\\
$^1$Department of Physics, Kyoto University, Kyoto 606-8502, Japan
}

\recdate{
January 30, 1999}

\abst{%
We show that neutrino masses with large mixing between $\nu_\mu$
and $\nu_\tau$ are naturally reproduced in a supersymmetric $E_6$
grand unification model with an anomalous $U(1)_X$ symmetry. We
propose a simple scenario which incorporates a novel mechanism called
`E-twisting' by which all the characteristic features of the fermion
mass matrices, not only the quark/lepton Dirac masses but also the
neutrino's Majorana masses, can be reproduced despite the fact that all the
members in {\bf27} of each generation are assigned a common $U(1)_X$
charge.

}
\begin{document}

\maketitle

\section{Introduction}
A recent report from Superkamiokande strongly indicates two important
facts\cite{kamioka} that have led us to unification groups
including left-right symmetry. The first fact is the extremely small
masses of the neutrinos. This is most naturally explained by the
existence of right-handed neutrinos with their Majorana masses of order
of $10^{13}$ GeV.\cite{seesaw} The existence of  right-handed
neutrinos implies that Nature prefers left-right symmetry, and we are led
to a left-right symmetric unification group like $SO(10)$. The second
fact is the large neutrino mixing angle, which is needed to explain the
zenith angle dependence of the observed atmospheric muon neutrino
spectrum. If one wishes to understand the mass matrices in unified
theories larger than $SU(5)$, this large neutrino mixing angle is 
apparently unnatural compared with the ordinary quark lepton mass
matrices. In $SO(10)$ GUT, for instance, quarks and leptons appear on
the same footing, and Dirac-type masses for the neutrinos and charged
leptons should be parallel to those for the up- and down-type quarks,
respectively. Thus the most natural prediction would be that the 
neutrino mixing is also very small with hierarchical mass structure 
among generations. This naive prediction of parallelism between quark 
and lepton sectors, however, was found to be maximally violated in the 
neutrino sector. Thus the neutrino masses and mixings seem to require 
something beyond even $SO(10)$.\footnote{ If the right-handed neutrino 
Majorana mass matrix $M_R$ can be chosen freely, then we can have any 
desired form for the left-handed neutrino Majorana mass matrix 
$M_\nu=M_D^{\rm T}M_R^{-1}M_D$, whatever Dirac mass matrix $M_D$ is used
for the neutrino in the GUT. But the problem is whether the form of 
$M_R$ can be derived in a given GUT framework.\cite{Majorana} In the 
scenarios which use the Frogatt-Nielsen mechanism, it seems that $M_R$ 
always has a hierarchical structure. }

Here we would like to remark also that the up and down quark mass 
matrices already violate the parallelism: The hierarchical structure of 
the up quark sector is far stronger than that of down quark sector. Also
it has long been one of the most important problems why the top quark is
far heavier than the bottom quark. 

If one wishes to explain the inter-generation hierarchical mass
structure, one natural possibility is the Frogatt-Nielsen
mechanism\cite{frogattnielsen} using an anomalous $U(1)_X$ charge.
However, if the unified gauge group is
$SO(10)$ or larger, all the fermion members in each generation have 
a common $U(1)_X$ charge quantum number, so that
the naive application of the
Frogatt-Nielsen mechanism predicts a completely parallel hierarchical
mass matrix for all the quarks and leptons, contradicting observations.
The naive prediction of this complete parallelism between quark and lepton
mass matrices is, however, a result of our prejudice that the three
generations of fermions are a mere repetition of the same structure,
namely, a parallel family structure. If the parallelism between families
is broken, then the parallelism between mass matrices of quarks and leptons
can also be avoided.
The importance of breaking parallel family structure
was first emphasized by Yanagida.\cite{yanagida} \
Then the question is how the unification model, and what unification group,
can avoid this parallel family structure.
Among the simple groups larger than
$SO(10)$,  $E_6$ is essentially the unique group that
admits complex representations in a manner consistent with the chiral
structure of our low-energy fermions. The exceptional group $E_6$ has
been investigated by many authors\cite{ref:E6,barbieri,KugoSato,%
ramond1,ramond2,ramond3,ramond4,joeyanagida,ref:matsuoka}
as an attractive unification group, because of its desirable
features: 1) it is automatically anomaly-free; 
2) all the basic fermions of one
generation belong to a single irreducible representation ${\bf 27}$; 
and  3) all the Higgs fields necessary for symmetry
breaking are supplied by the fermion bilinears.
On the other hand, 10 dimensional $E_8\times E_8$ heterotic string theory
has been thought to be very attractive as a unified theory including gravity
from which the low energy $N=1$ supersymmetric standard model may be
derived.\cite{string}
The Calabi-Yau compactification into 4 dimensions may naturally
produce $E_6$ gauge symmetry. In this kind of string model, there
exists an anomalous $U(1)$ which is to be cancelled by the so-called
Green-Schwarz mechanism.\cite{ramond4}
In supersymmetric model this anomalous $U(1)$
was found to play an essential role for explaining the hierarchy of
fermion masses via Frogatt-Nielsen mechanism.\cite{ibanetzross}

The purpose of this paper is to show that there is a novel and natural
mechanism in a supersymmetric $E_6$ grand unification model with an
anomalous $U(1)_X$ symmetry, with which the parallel family structure
can be avoided and all the characteristic features
of the fermion mass matrices (not only the quark/lepton Dirac masses
but also the neutrino's Majorana masses) can be reproduced.
We propose and examine a very simple scenario which incorporates this
novel mechanism called `E-twisting'.

In \S2 we present the basic framework of our $E_6$ model and
point out that there is a freedom of an $SU(2)_E$ inner automorphism
in $E_6$ in embedding $SO(10)$ such that $SU_{\rm GG}(5)\subset SO(10)\subset
E_6$.
As a reflection of this $SU(2)_E$ symmetry, the fundamental
representation {\bf27} contains two $SU(5)$ ${\bf5}^*$ and two ${\bf1}$
components, both giving spinor representations of $SU(2)_E$.
This $SU(2)_E$ transformation plays an important role in explaining the
large mixing angle of neutrinos while keeping very small family
mixings in the quark sector. In \S3 we determine the
$U(1)_X$ charges assigned to the three generation matter superfields
$\Psi_i({\bf27})$ belonging to the {\bf27} representations of $E_6$.
In \S4 we discuss ${\bf 5}^*$ family structure; namely, we consider 
which three ${\bf 5}^*$ components become the usual down quarks and 
charged leptons among the six ${\bf 5}^*$ components in the three 
{\bf27} representations. Describing three typical options, parallel 
family structure, nonparallel family structure, and E-twisted structure,
we argue that the E-twisted structure gives the simplest option which 
can reproduce the down quark spectrum and the Cabibbo-Kobayashi-Maskawa 
(CKM) mixings, and, at the same time, the large lepton mixing angle. In 
\S5 we examine the right-handed neutrino Majorana masses which naturally
come from higher dimensional operators in the present scenario. Section 
6 is devoted to the summary and discussion.

\section{$\mbf E_6$ unification model and E-symmetry}

We consider here the simplest example of a supersymmetric $E_6$
unification model which reproduces the large neutrino mixing angle as
well as the quark and lepton masses and mixings.

Aside from the $E_6$ gauge vector multiplet, we introduce several chiral
multiplets which belong to the fundamental representation ${\bf 27}$ or
${\bf 27^*}$, or singlet ${\bf1}$,  and carry a quantum number (denoted by
$X$) of anomalous $U(1)_X$.
\begin{enumerate}
\item  The quarks and the leptons  of the three generations
are included in three chiral multiplets $\Psi_i$ $(i=1,2,3)$ of the ${\bf27}$
representation, to which we assign odd R-parity and $U(1)_X$ charges
$X=\ff_i$.  (The values $\ff_i$ will be determined later.)

\item
We introduce two pairs of Higgs fields, $(H^h, \bar
H^{-h})$  and $(\Phi^\QQ$, $\bar \Phi^{-\QQ})$,
of ${\bf27}$ and ${\bf27^*}$ chiral multiplets,
which are R-parity even and carry
the $U(1)_X$ charges $X=\pm h$ and $X=\pm\QQ$, respectively.
Their neutral components develop vacuum expectation values (VEVs) and
give superheavy masses to the fermions other than the usual low-energy
quarks and leptons and also the light masses to  the ordinary fermions.

\item In addition to these, we introduce an $E_6$ singlet field $\Theta$
with $X=-1$ and R-parity even that  supplements the Yukawa couplings so
as to match the $U(1)_X$ charge.
\end{enumerate}
Then the R-parity, $U(1)_X$ and $E_6$ invariant superpotentials giving Yukawa
interactions are
\begin{eqnarray}
   W_Y(H) &=&
        y\,\Psi_i({\bf 27})\Psi_j({\bf 27})H({\bf27})
        \left({\Theta\over M_P}\right)^{\ff_i+\ff_j+h},           \nn
   W_Y(\Phi) &=&
        y'\,\Psi_i({\bf 27})\Psi_j({\bf 27})\Phi({\bf27})
        \left({\Theta\over M_P}\right)^{\ff_i+\ff_j+\QQ}.
\label{yukawa}
\end{eqnarray}
The coupling constants $y$ and $y'$ may generally depend on $(i,j)$,
but we assume they are all of order 1, and so we can suppress the $(i,j)$
dependence in the present order of magnitude discussion.
The  hierarchical mass structure of fermions
are explained via the Frogatt-Nielsen mechanism; that is,
the powers of the $E_6$ singlet field $\Theta$
in this higher dimensional Yukawa coupling (\ref{yukawa}) are required
by $U(1)_X$ quantum number matching,  and
the effective Yukawa couplings are suppressed by
corresponding powers of $\lambda=\langle\Theta\rangle/M_P$.
We assume $\lambda=\langle\Theta\rangle/M_P\sim0.22$ henceforth.

Each fundamental representation $\Psi({\bf 27})$ is decomposed
under $SO(10)\times U(1)_\VT\subset E_6$ as
\begin{equation}
\begin{array}{ccccccccl}
 {\bf 27} & = &  {\bf 16}_1 &+& {\bf 10}_{-2} &+& {\bf 1}_4\ , 
&\quad & (\alpha=1,2,\cdots,16)   \\
 \Psi_A:     &  &  \psi_\alpha&& \psi_M && \psi_0
&\quad & (M=1,2,\cdots,10)    \\
\end{array}
\label{eq:so1027}
\end{equation}
and under $SU(5)\times U(1)_V\subset SO(10)$ as
\begin{equation}
\begin{array}{ccccccc}
 {\bf 16} & = &  {\bf 10}_{-1} &+& {\bf 5}^*_{3} &+& {\bf 1}_{-5}\ ,     \\
 \psi_\alpha:     &  &  \left[u^{ci}, \pmatrix{u_i\cr d_i\cr}, e^c\right] &&
(d^{ci}, e, -\nu) && \nu^c       \\
\end{array}
\label{eq:su516}
\end{equation}
\begin{equation}
\begin{array}{ccccc}
 {\bf 10} & = &  {\bf 5}_{2} &+& {\bf 5}^*_{-2}\ ,     \\
 \psi_M:     &  &  \pmatrix{D_i\cr E^c\cr -N^c\cr} &&
(D^{ci}, E, -N)  \\
\end{array}
\label{eq:su510}
\end{equation}
\begin{eqnarray}
 {\bf 1} & = &  {\bf 1}_0\ . \nn
   \psi_0:  &  &  S
\label{eq:su1}
\end{eqnarray}
Note that the representations ${\bf5^*}$ and ${\bf1}$ of $SU(5)$ appear
twice here, while ${\bf10}$ and ${\bf5}$ appear only once.
This suggests that we can define a parity-like transformation,
which we call `E-parity':
\begin{equation}
\hbox{E-parity transformation:}\qquad
\matrix{
\Psi({\bf16},{\bf5^*}) &\leftrightarrow & \Psi({\bf10},{\bf5^*})\ , \cr
\Psi({\bf16},{\bf1})   &\leftrightarrow & \Psi({\bf1},{\bf1})\ .\cr}
\end{equation}
The other $SU(5)$ components, $\Psi({\bf16},{\bf10})$ and $\Psi
({\bf10},{\bf5})$,  remain intact.
Actually, this E-parity turns out to be a $\pi$ rotation of a certain
$SU(2)\subset E_6$: Indeed, even if we fix the $SU(5)$ subgroup in $E_6$ as the
usual $SU(5)_{\rm GG}$ of Georgi-Glashow, the embedding of $SO(10)$ into
$E_6$ such that $SU(5)_{\rm GG}\subset SO(10)\subset E_6$  is not unique, but
possesses a freedom of rotation of 
$SU(2)$.\footnote{The so-called ``flipped $SU(5)$'' gives another example
similar to this situation. There, the embedding of $SU(5)$ into $SO(10)$
such that $SU(3)_c\times SU(2)_L\subset SU(5)\subset SO(10)$ holds, is
not unique but possesses a freedom of rotation of  $SU(2)$.
This $SU(2)$ is in fact the usual $SU(2)_R$ of
$SU(4)_{{\rm Pati}\hbox{-}{\rm Salam}}\times SU(2)_L\times SU(2)_R\subset
SO(10)$.
The $SU(5)$ group of the
flipped $SU(5)$ model is given from the usual $SU(5)_{\rm GG}$ by a $\pi$
rotation of $SU(2)_R$. This can be understood by an argument  very similar
 to that given in the next paragraph, by
considering the decomposition of $SO(10)$ {\bf16} under the subgroups
$SU(3)_c\times SU(2)_L\times SU(2)_R\subset SU(4)\times SU(2)_L\times SU(2)_R$;
${\bf16} =
({\bf4},\,{\bf2}_L,\,{\bf1}_R) + ({\bf4}^*,\,{\bf1}_L,\,{\bf2}_R) =
({\bf3}_c,\,{\bf2}_L,\,{\bf1}_R) + ({\bf1}_c,\,{\bf2}_L,\,{\bf1}_R) +
({\bf3}_c^*,\,{\bf1}_L,\,{\bf2}_R) + ({\bf1}_c,\,{\bf1}_L,\,{\bf2}_R)$.}

To see this fact it would be
easiest if we consider another maximal
subgroup $SU(6)\times SU(2)_E\subset E_6$,  where the former $SU(6)$ contains
$SU(5)_{\rm GG}\times U(1)_Z$ such that the first 5 entries of the
fundamental representation ${\bf6}$ give  the fundamental representation
${\bf5}$ of $SU(5)_{\rm GG}$. Under this $SU(5)_{\rm GG}\times SU(2)_E\times
U(1)_Z
\subset SU(6)\times SU(2)_E\subset E_6$, the ${\bf 27}$
is decomposed into
\begin{eqnarray}
{\bf 27} &=&
\underbrace{({\bf5^*},{\bf2})_{-1} + ({\bf1},{\bf2})_{5}}
_{({\bf6^*},\, {\bf2})}
+ \underbrace{({\bf10},{\bf1})_2 + ({\bf5},{\bf1})_{-4}}
_{({\bf15},\,{\bf1})}\ .
\label{eq:su6decomp}
\end{eqnarray}
This decomposition of ${\bf27}$ clearly shows that the two
${\bf5^*}$ as well as ${\bf1}$ belong to a doublet of $SU(2)_E$,
and hence they are rotated into each other as a spinor by this $SU(2)_E$.
In view of the $U(1)_\VT$ and $U(1)_V$ charges of ${\bf 5^*}$ appearing
in Eqs.~(\ref{eq:so1027}) --- (\ref{eq:su510}) and
the $U(1)_Z$ charge here,
we can see that the following identifications are possible
between these $U(1)$ charges and the third component $E_3$ of $SU(2)_E$
generators $E_j$ ($j=1,2,3$):
\begin{equation}
V=5E_3 - {1\over2}Z \ , \qquad
\VT=3E_3 + {1\over2}Z \ .
\end{equation}
Collecting the representation components carrying the same $\VT$ charge,
we easily find that the $SO(10)$ multiplets ${\bf16}, {\bf10}$ and ${\bf1}$
are formed as follows:
\begin{eqnarray}
{\bf16}_1&=& ({\bf5^*},E_3=+1/2)_{-1} + ({\bf1},E_3=-1/2)_{5} +
({\bf10},{\bf1})_2\ , \nn
{\bf10}_{-2}&=& ({\bf5^*},E_3=-1/2)_{-1} + ({\bf5},{\bf1})_{-4}\ , \nn
{\bf1}_4&=& ({\bf1},E_3=+1/2)_{5}\ .
\label{eq:2.9}
\end{eqnarray}
But this identification of $SO(10)$ must be possible even after
performing an arbitrary $SU(2)_E$ rotation
$\exp(i\theta^jE_j)$. Then the $SO(10)$ generators $T_{MN}({\mbf\theta})$ for
the case of an 
angle ${\mbf\theta}$ are given by $\exp(i{\mbf\theta}\ncdot {\mbf
E})T_{MN}\exp(-i{\mbf\theta}\ncdot {\mbf E})$ in terms of the original
$SO(10)$ generators $T_{MN}$. This $\exp(i{\mbf\theta}\ncdot {\mbf E})$
defines an inner automorphism of $E_6$ which gives a three parameter
family of embeddings of $SO(10)$ into $E_6$ such that $SU(5)_{\rm GG}\subset
SO(10)\subset E_6$ holds. We may call this $\exp(i{\mbf\theta}\ncdot {\mbf E})$
E-symmetry,  and the E-parity transformation introduced above is merely
a special case of
$\theta=\pi$, $\exp(i\pi E_1)$ (or $\exp(i\pi E_2)$).\footnote{%
It may worth noting that this E-symmetry group $SU(2)_E$ is identical with
an $SU(2)$ subgroup of $SU(3)_R$ which appears in
the famous maximal subgroup $SU(3)_c\times SU(3)_L\times SU(3)_R$ of $E_6$.
But $SU(2)_E$ is the $SU(2)$ subgroup of $SU(3)_R$ acting on the
second and third entries of ${\bf3}$ of $SU(3)_R$, while the
usual $SU(2)_R$ of left-right symmetric theory is another
$SU(2)$ subgroup of $SU(3)_R$ acting on the
first and second entries of ${\bf3}$. See the Appendix for details.}
Indeed, with this $\exp(i\pi E_1)$ rotation, the $E_3$ generator is
rotated into $E'_3= -E_3$,  and hence the $E_3=\pm1/2$ eigenstates of
the spinor representation are interchanged, implying that the two
${\bf5^*}$ states in ${\bf16}$ and ${\bf10}$, as well as the two
${\bf1}$ states in ${\bf16}$ and ${\bf1}$, in Eq.~(\ref{eq:2.9}) are
interchanged.

In Eq.~(\ref{yukawa}) the Yukawa couplings for individual
components are determined by the $E_6$
group relation: The $E_6$ invariant trilinear
in ${\bf 27}$ is given by a totally symmetric tensor $\Gamma^{ABC}$ 
in the form\cite{KugoSato}
\begin{eqnarray}
\Gamma^{ABC}\Psi_{1A}\Psi_{2B}\Psi_{3C} &=& \psi_{1M}\psi_{2M}\psi_{30} +
\psi_{1\alpha}^{\rm
T}\big(\psi_{2M}{C\sigma_M\over\sqrt2}\big)^{\alpha\beta}\psi_{3\beta} \nn
&& {}+ [\ \hbox{cyclic permutations of } (1,2,3)\ ]\ ,
\label{yukawacomp}
\end{eqnarray}
where $\psi_{jM}$, $\psi_{j\alpha}$ and $\psi_{j0}$ are the 
$SO(10)$ vector, spinor and singlet components of $\Psi_j$ defined in
Eq.~(\ref{eq:so1027}), and $\sigma_M$ and $C$ are the 
$16\times16$ $\gamma$-matrix
and the charge conjugation matrix of $SO(10)$, respectively. (See the 
Appendix for a more explicit component expression of 
Eq.~(\ref{yukawacomp}).) For our present purpose of discussing the 
fermion masses and mixings coming from the VEVs of the Higgs fields, we 
list here the explicit forms for the terms in 
$\Gamma^{ABC}\Psi_{1A}\Psi_{2B}H_C$ containing the neutral Higgs 
components $H({\bf 1}), H({\bf 16,1}), H({\bf 16,5}^*), H({\bf10,5^*})$ 
and $H({\bf10,5})$, which correspond to the components $S,\ \nu^c,\ 
-\nu,\ -N$ and $-N^c$, respectively, of $\Psi({\bf27})$ in 
Eqs.~(\ref{eq:so1027}) -- (\ref{eq:su510}):
\begin{equation}
\begin{array}{cccccc}
H({\bf 1}) &\Bigl[&
(D^{ci}\ E\ -N)_1\pmatrix{D_i\cr E^c\cr -N^c\cr}_2
&+&(1\leftrightarrow2)
&\Bigr]\ , \\
&& \hbox{\footnotesize\bf10(5$^*$)}\times\hbox{\footnotesize\bf10(5)} &&
&
\end{array}
\label{singlet}
\end{equation}
\begin{equation}
\begin{array}{cccccc}
H({\bf 16,1}) &\Bigl[&
-(D_i\ E^c\ -N^c)_1\pmatrix{d^{ci}\cr e\cr -\nu\cr}_2
&+&(1\leftrightarrow2)&\Bigr]\ , \\
&& \hbox{\footnotesize\bf10(5)}\times\hbox{\footnotesize\bf16(5$^*$)} && &
\end{array}
\label{singlet16}
\end{equation}
\begin{equation}
\begin{array}{cccccccc}
H({\bf 16,5}^*) &\bigl[&
-\nu^c_1N^c_2 &+& (\ \ D^{ci}_1d_{i2}\ + \ E_1e^c_2\ \ )
&+&(1\leftrightarrow2) &\bigr]\ , \\
&&\hbox{\footnotesize\bf16(1)}\times\hbox{\footnotesize\bf10(5)}
&&\hbox{\footnotesize\bf10(5$^*$)}\times\hbox{\footnotesize\bf16(10)}&&&
\end{array}
\label{downmass16}
\end{equation}
\begin{equation}
\begin{array}{cccccccc}
H({\bf 10,5}^*) &\bigl[&
S_1N^c_2
&-& (\ \ d^{ci}_1d_{i2}
\ + \  e_1e^c_2\ \ )
&+&(1\leftrightarrow2)&\bigr]\ , \\
&&\hbox{\footnotesize\bf1}\times\hbox{\footnotesize\bf10(5)}
&&\hbox{\footnotesize\bf16(5$^*$)}\times\hbox{\footnotesize\bf16(10)}&&&
\end{array}
\label{downmass10}
\end{equation}
\begin{equation}
\begin{array}{cccccccc}
H({\bf10,5}) &\bigl[&
S_1N_2
&-& (\ \ u^{ci}_1u_{i2}
\ \ \ +\ \ \ \nu_1\nu^c_2\ \ )
&+&(1\leftrightarrow2)&\bigr]\ . \\
&&\hbox{\footnotesize\bf1}\times\hbox{\footnotesize\bf10(5$^*$)}
&&\hbox{\footnotesize\bf16(10)}{\times}\hbox{\footnotesize\bf16(10)}
\ \ \hbox{\footnotesize\bf16(5$^*$)}{\times}\hbox{\footnotesize\bf16(1)}&&&
\end{array}
\label{upmass}
\end{equation}

\section{Determination of $\mbf\mbfplus U(1)_X$ charges}

In ${\bf 27}$, $SU(5)$ ${\bf 10}$ and ${\bf5}$ components appear once,
and hence 
the mass matrix of the up quark sector uniquely comes from
the term $\Psi({\bf 16,10})\Psi({\bf 16,10})H({\bf 10,5})$ in
Eq.~(\ref{upmass}), which always yields a symmetric matrix. Let
$H({\bf 10, 5})$ develop  a VEV, $\langle H({\bf 10, 5})\rangle=v_u$, giving
the up
sector masses. We can take the $U(1)_X$ charge of the Higgs fields
$H$ to be $h=0$ without loss of generality. With this convention, we now
determine the $U(1)_X$ charges $\ff_i$ for the three families of
$\Psi_i({\bf27})$.

Experimentally, the up sector mass matrix is known to take the form
\begin{eqnarray}
M_u=m_t \times\bordermatrix{
                   &{\bf 10}_1     & {\bf 10}_2   &{\bf 10}_3 \cr
   {\bf 10}_1      &   \lambda^{7}      &   (\lambda^5)     & (\lambda^3)
\cr
   {\bf 10}_2      &   (\lambda^5)      &   \lambda^4       &  (\lambda^2)
\cr
   {\bf 10}_3      &   (\lambda^3)      &   (\lambda^2)     &  1      \cr}\ ,
\label{expupmass}
\end{eqnarray}
where $\lambda\sim0.22$ is of the order of the Cabibbo angle,  and
the brackets indicate that the corresponding matrix element may be
less than the order indicated, since the CKM angles
represent the difference between the up and down sectors.
We take $\ff_3=-h/2$, which is zero by our convention for $h$. Then,
in view of the diagonal 33, 22 and 11 elements of this mass matrix,
we see that we must take $\ff_2=2$,  and
$\ff_1$ should be either 3 or 4.

In order to fix the $U(1)_X$ charge to either 3 or 4,
we use information concerning the Cabibbo angle. But to discuss
the Cabibbo angle, we need  information regarding  the down sector mass
matrix. The mass matrix of the down sector has, however, a complication
that it comes from the Yukawa couplings of
two types, Eqs.~(\ref{downmass10}) and (\ref{downmass16})
 owing to the mixing of the two ${\bf5^*}$ representations in ${\bf27}$
both for the right-handed down quark $d^c$ and the Higgs $H$. 
We discuss this problem in detail in the next section.
The Cabibbo angle is given by
$M_{u\,12}/M_{u\,22}-M_{d\,12}/M_{d\,22}$, 
where  $M_u$ and $M_d$ are the mass matrices of the up and down quark sectors.
We will see 
that, in the scenario discussed there,
$M_{d\,12}/M_{d\,22}$ as well as $M_{u\,12}/M_{u\,22}$
is of the order of $\lambda^{\ff_1-\ff_2}$, determined by the $U(1)_X$ charge
assignment of the first and second families.
Thus  the Cabibbo angle of order $\lambda$ requires $\ff_1-\ff_2=1$,
from which we must uniquely choose $\ff_1=3$.

Now we have fixed the $U(1)_X$ charge as
\begin{equation}
      X(H)=0\ , \qquad 
    \big(X(\Psi_1),\,X(\Psi_2),\,X(\Psi_3)\big)=(3,\,2,\,0)\ ,
\label{fcharge}
\end{equation}
from which we have
\begin{eqnarray}
M_u=m_t\times\bordermatrix{
                     &{\bf 10}_1  & {\bf 10}_2   &{\bf 10}_3 \cr
   {\bf 10}_1        &   \lambda^6     &   \lambda^5       &  \lambda^3   \cr
   {\bf 10}_2        &   \lambda^5     &   \lambda^4       &  \lambda^2   \cr
   {\bf 10}_3        &   \lambda^3     &   \lambda^2       &   1     \cr}\ ,
\label{theoryupmass}
\end{eqnarray}
with $m_t\equiv yv_u$. The $u$ quark mass of order $\lambda^6$ is a
bit large, so 
that a small cancellation should occur in the computation of the
correct eigenvalue for $u$.

\section{ ${\bf 5}^*$ family structure}

Now that we have determined the $U(1)_X$ charges of the three families
$\Psi_i({\bf27})$, our next task is to look for a possible ${\bf 5}^*$
family structure which can reproduce the down quark spectrum and CKM
mixings, and, at the same time, the large lepton mixing angle.

As mentioned above, there are two ${\bf5^*}$ components in each
${\bf27}$. The Higgs doublet, which survives down to the low energy 
region, can in general be an arbitrary mixed state of the two ${\bf 
5}^*$ components in $H({\bf27})$:
\begin{equation}
H({\bf 5^*})=H({\bf 10,5^*})\cos\theta+H({\bf 16,5^*})\sin\theta \ .
\label{eq:higgs}
\end{equation}
This is just an $H({\bf 10,5^*})$ of the E-rotated $SO(10)$ by angle
$2\theta$,
corresponding to generators $T_{MN}(2\theta)=\exp(2i\theta
E_2)T_{MN}\exp(-2i\theta E_2)$.
Of course, it should be noted that this angle $\theta$ itself has no physical
meaning, but becomes meaningful once the other down quarks or Higgs
specify the direction of the $SO(10)$. In the conventional treatment,
the low energy Higgs doublet is identified with $H({\bf 10,5^*})$, 
corresponding to the choice $\cos\theta=1$, which implies that the $SO(10)$
direction is specified by the Higgs field. Here, in our scheme, the $SO(10)$
direction will be determined by the down quark sector.

Let us proceed to the three families of (right-handed) down quarks. 
They can in general be any three linear combinations of the six
${\bf5^*}$ components in the three $\Psi_i({\bf27})$.
We here consider three typical scenarios.
\begin{enumerate}
\item  Parallel family structure \\
 Each family of low energy fermions ${\bf5^*}$ and ${\bf 10}$ is 
contained in a ${\bf 16}$ representation of (a fixed) $SO(10)$: 
\begin{equation}
({\bf 5}^*_1,\ {\bf 5}^*_2,\ {\bf 5}^*_3)
= \bigl(\Psi_1({\bf 16,5^*}),\ \Psi_2({\bf 16,5^*}),\ \Psi_3({\bf
16,5^*})\bigr)\ .
\label{parallel}
\end{equation}
\item Nonparallel family structure \\
Inter-generation mixings are so large that
the three ${\bf 5^*}$ components do not come one from each  
$\Psi_1$, $\Psi_2$ and $\Psi_3$. 
Such an example is
\begin{equation}
({\bf 5}^*_1,\ {\bf 5}^*_2,\ {\bf 5}^*_3)
= \bigl(\Psi_1({\bf 16,5^*}),\ \Psi_2({\bf 16,5^*}),\ \Psi_2({\bf
10,5^*})\bigr)\ .
\label{nonparallel}
\end{equation}
A structure similar to this was first proposed by
Yanagida.\cite{yanagida,joeyanagida,NomuraYanagida}
\item E-twisted structure \\
Some of the $\Psi_i({\bf 16,5^*})$ are replaced by $\Psi_i({\bf 10,5^*})$ 
in the parallel family structure (\ref{parallel}). For example the 
E-twisted version of the third family is given by
\begin{equation}
({\bf 5}^*_1,\ {\bf 5}^*_2,\ {\bf 5}^*_3)
= \bigl(\Psi_1({\bf 16,5^*}),\ \Psi_2({\bf 16,5^*}),\ \Psi_3({\bf
10,5^*})\bigr)\ .
\label{twisting}
\end{equation}
This structure implies that the third family falls into ${\bf16}$ of
an $SO(10)$ E-twisted from that of the other two
families.\footnote{In terms of the component notation in
Eqs.~(\ref{eq:su516}) and (\ref{eq:su510}), our E-twisted scenario is to
take $(d^c_1,\,d^c_2,\,D_3^c)$ for down quarks and $(e_1,\,e_2,\,E_3)$
for leptons. Somewhat similar twisting in the down quark and lepton
sectors was  considered by Haba et al.\cite{ref:matsuoka} in the
$SU(6)'\times SU(2)_R$ unified theory, in which they took essentially
$(O(1)d^c_1+O(1)d^c_2,\,D_1^c,\,D_3^c)$
and $(e_1,\, E_1,\, O(1)E_2+O(1)E_3)$.}
\end{enumerate}

\noindent
Among these three possibilities, it is obvious that the first option
(the parallel family structure) predicts completely parallel mass
matrices for all the fermions, the up and down quarks, leptons and
neutrinos. Thus it can reproduce neither the down quark masses nor the
large neutrino mixing angle.

As for the second option (nonparallel family structure), it turns out
that we need a higher representation Higgs field in order to make
the other fermion components superheavy, as is seen as follows. 
We need to make
$\Psi_1({\bf10,5^*})$, $\Psi_3({\bf16,5^*})$ and $\Psi_3({\bf10,5^*})$,
as well as all the ${\bf5}$ components, $\Psi_i({\bf10,5})$,
superheavy. This could be most economically achieved as shown
in Fig.~1; namely,
\begin{figure}[tb]
   \epsfxsize= 6.5cm
   \centerline{\epsfbox{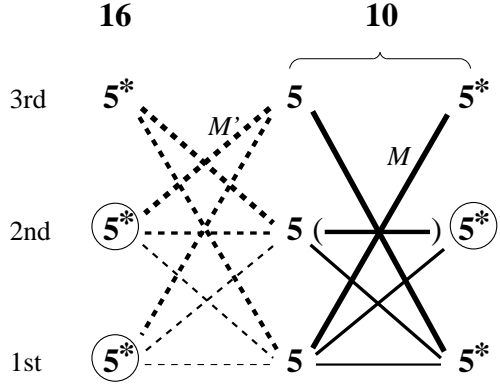}}
   \caption{Nonparallel family structure in which
the ${\bf5}^*$ components enclosed by circles are the dominant components
which remain light.
The solid and dotted lines denote heavy mass terms given by
$\VEV{\Phi^{\QQ=-3}({\bf1})}$
and $\VEV{\Phi^{\QQ=-2}({\bf 16,1})}$, respectively.
The $U(1)_X$-suppressed terms with higher powers of
$\VEV{\Theta}/M_P\sim\lambda$
are indicated by thinner lines.
The bracketed line connecting ${\bf5}$ to ${\bf5}^*$ in the
2nd generation {\bf10} should be absent for this scenario to work,
which can be realized if $\Phi^{\QQ=-3}$ is the {\bf351} coupling to
$({\bf27}\times{\bf27})_A$ anti-symmetrically.
}
\end{figure}
we first introduce a Higgs field $\Phi^{\QQ=-3}$ with
the $U(1)_X$ charge $-3$ such that its VEV
$\langle\Phi^{\QQ=-3}({\bf 1})\rangle=M/y'$ in the $SO(10)$ singlet component
can yield mass terms
$M\Psi_3({\bf10,5^*})\Psi_1({\bf10,5})$ and
$M\Psi_1({\bf10,5^*})\Psi_3({\bf10,5})$
via the Yukawa coupling in Eqs.~(\ref{yukawa}) and (\ref{singlet}).
This VEV also yields other mass terms by the Yukawa interactions
(\ref{yukawa}),  which are generally allowed by supplementing powers of
$(\Theta/M_P)$ to match the $U(1)_X$ quantum number:
\begin{eqnarray}
M\times
\bordermatrix{
  &\Psi_1({\bf 10,5^*})&\Psi_2({\bf 10,5^*})&\Psi_3({\bf 10,5^*})\cr
\Psi_1({\bf 10,5})   & \lambda^4         &   \lambda^2            & 1  \cr
\Psi_2({\bf 10,5})   & \lambda^2         &   (\lambda)            & 0  \cr
\Psi_3({\bf 10,5})   &  1           &    0              & 0  \cr }\ .
\end{eqnarray}
The 22 matrix element is bracketed here for the reason
that becomes clear shortly. Next we add another Higgs field $\Phi
^{\QQ=-2}$ with  $U(1)_X$ charge $-2$ such that its VEV $\langle\Phi
^{\QQ=-2}({\bf 16,1})\rangle=M'/y'$ in the $SO(10)$ ${\bf16}$ $SU(5)$ singlet
component can give the mass term $M'\Psi_3({\bf16,5^*})\Psi_2({\bf10,5})$ via
the coupling (\ref{singlet16}). Of course, this coupling simultaneously
yields an (unwanted) mass term $M'\Psi_2({\bf16,5^*})\Psi_3({\bf10,5})$ with
the same strength $M'$. Including also the subdominant terms suppressed
by powers of $\Theta/M_P$, this VEV gives the mass matrix
\begin{eqnarray}
M'\times
\bordermatrix{
      &\Psi_1({\bf 16,5^*})&\Psi_2({\bf 16,5^*})&\Psi_3({\bf 16,5^*}) \cr
\Psi_1({\bf 10,5}) & \lambda^4          &   \lambda^2           & \lambda  \cr
\Psi_2({\bf 10,5}) & \lambda^2          &   \lambda^2           &  1   \cr
\Psi_3({\bf 10,5}) &  \lambda          &    1             &  0   \cr
}\ .
\end{eqnarray}
Suppose that $M\gg M'$; namely the breaking scale of $E_6$ down to $SO(10)$
is higher than the breaking scale of $SO(10)$ down to $SU(5)$.
Then, first the pairs
$\Psi_3({\bf10,5^*})$ and $\Psi_1({\bf10,5})$, and
$\Psi_1({\bf10,5^*})$ and $\Psi_3({\bf10,5})$ acquire a superheavy mass
$M$ and decouple from the others. Next,  we expect
that the pair consisting of $\Psi_3({\bf16,5^*})$ and
$\Psi_2({\bf10,5})$ becomes
superheavy by the VEV $\langle\Phi^{\QQ=-2}({\bf 16,1})\rangle=M'/y'$. However,
for
the component $\Psi_2({\bf10,5})$, there is also another mass term $\lambda
M\Psi
_2({\bf10,5})\Psi_2({\bf10,5^*})$ with strength $\lambda M$ coming from the
coupling to the VEV $\langle\Phi^{\QQ=-3}({\bf 1})\rangle=M/y'$. In order to
keep $\Psi_2({\bf10,5^*})$ light while making the pair 
$\Psi_3({\bf16,5^*})$ and $\Psi_2({\bf10,5})$ superheavy, the condition 
$M'\gg\lambda M$ should hold, but this is incompatible with the first 
assumption, $M\gg M'$. (Note that, without the condition $M\gg M'$, the 
presence of the mass term $M'\Psi _2({\bf16,5^*})\Psi_3({\bf10,5})$ 
would have caused a large mixing between $\Psi_2({\bf16,5^*})$ and 
$\Psi_1({\bf10,5^*})$, contrary to our desire to keep the component 
$\Psi_2({\bf16,5^*})$ light.) The only way to avoid this situation is 
therefore to forbid the problematic mass term $\lambda 
M\Psi_2({\bf10,5})\Psi_2({\bf10,5^*})$, which would be realized if the 
first Higgs field $\Phi^{\QQ=-3}$ with $\langle\Phi^{\QQ=-3}({\bf 
1})\rangle=M/y'$ were not a ${\bf 27}$ but rather ${\bf 351}$ 
representation of $E_6$ whose Yukawa couplings to $\Psi_i({\bf27})
\Psi_j({\bf27})$ are only antisymmetric in $(i,j)$.

The simplest and most attractive option is thus the third one,
E-twisted structure,
which we investigate in the following sections.

\section{Down sector masses in E-twisted structure}

Let us investigate the case of E-twisted structure.
In this case, we need only one Higgs field, 
 $\Phi^{\QQ=-4}$, in addition to the usual
Higgs $H$,  and we suppose that they develop the following VEVs:
\begin{equation}
\langle\Phi^{\QQ=-4}({\bf 1})=S_\Phi\rangle=M/y'\ , \qquad
\langle H({\bf 16,1})=\nu^c_H\rangle=M'/y \ .
\label{singletvev}
\end{equation}
These give the following superheavy mass terms via the Yukawa coupling
(\ref{yukawa}) with Eqs.~(\ref{singlet}) and (\ref{singlet16})
(See fig.~2):
\begin{figure}[t]
   \epsfxsize= 6.5cm
   \centerline{\epsfbox{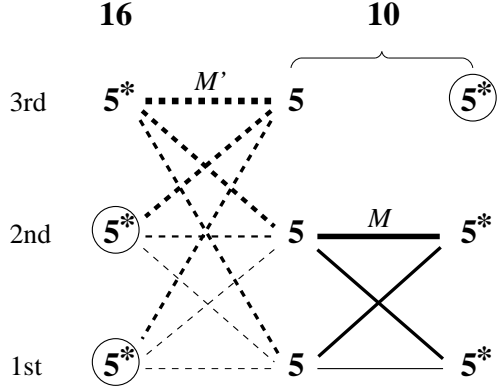}}
   \caption{E-twisted structure in which
the ${\bf5}^*$ components enclosed by circles are the dominant components
that remain light.
The solid and dotted lines denote heavy mass terms given by
$\VEV{\Phi^{\QQ=-4}({\bf1})}$
and $\VEV{H({\bf 16,1})}$, respectively.
The $U(1)_X$-suppressed terms with higher powers of
$\VEV{\Theta}/M_P\sim\lambda$
are indicated by thinner lines, as in Fig.~1.}
\end{figure}
\begin{equation}
\bordermatrix{
   &\!\Psi_1({\bf 16,5^*})\!\!&\!\!\Psi_2({\bf 16,5^*})\!\!&\!\!\Psi_3({\bf
16,5^*})\!\!
      &\!\!\Psi_1({\bf 10,5^*})\!\!&\!\!\Psi_2({\bf
10,5^*})\!\!&\!\!\Psi_3({\bf 10,5^*})\! \cr
\Psi_1({\bf 10,5})\!  & \lambda^6M' & \lambda^5M' & \lambda^3M' & \lambda^2M &
\lambda M  & 0    \cr
\Psi_2({\bf 10,5})\!  & \lambda^5M' & \lambda^4M' & \lambda^2M' &  \lambda M  &
 M   & 0    \cr
\Psi_3({\bf 10,5})\!  & \lambda^3M' & \lambda^2M' &   M'   &   0   &  0   & 0
 \cr }\ .
\label{masseex}
\end{equation}
Let us again assume that $M\gg M'$; i.e.,  the breaking scale of
 $E_6$ down to $SO(10)$ is higher than the breaking scale of $SO(10)$
 down to $SU(5)$. This case has a very simple structure in which $\langle\Phi
 ^{\QQ=-4}({\bf 1})\rangle=M/y'$ gives superheavy masses to $\Psi_i({\bf
 10,5^*})$ and $\Psi_i({\bf 10,5})$ with $i=1$ and $2$, while $\langle H({\bf
 16,1})\rangle=M'/y$ gives a slightly smaller superheavy mass to
 $\Psi_3({\bf 16,5^*})$ and $\Psi_3({\bf 10,5})$.
This now leaves us with the following three light (massless at this stage)
eigenstates:
\begin{eqnarray}
{\bf5}_1^* &=& \Psi_1({\bf 16,5^*}) + O(\lambda^3)\Psi_3({\bf
16,5^*})\  ,\nn
{\bf5}_2^* &=& \Psi_2({\bf 16,5^*}) + O(\lambda^2)\Psi_3({\bf
16,5^*})\ , \nn
{\bf5}_3^* &=& \Psi_3({\bf 10,5^*})\ ,
\label{eq:eigen5star}
\end{eqnarray}
up to unimportant minor mixings with heavier components. These three
states are essentially
$\bigl(\Psi_1({\bf 16,5^*}),\ \Psi_2({\bf 16,5^*}),\ \Psi_3({\bf
10,5^*})\bigr)$, 
as desired. Then, taking the Higgs mixing (\ref{eq:higgs}) into account,
the Yukawa couplings Eqs.~(\ref{downmass16}) and (\ref{downmass10}) lead
to the following form of the  down sector mass matrix:
\begin{eqnarray}
M_d^{\rm T}= yv_d \times
\bordermatrix{
             &  {\bf 5}^*_1  &   {\bf 5}^*_2  &  {\bf 5}^*_3 \cr
{\bf 10}_1   &  \lambda^6\cos\theta  &   \lambda^5\cos\theta  &
\lambda^3\sin\theta \cr
{\bf 10}_2   &  \lambda^5\cos\theta  &   \lambda^4\cos\theta  &
\lambda^2\sin\theta \cr
{\bf 10}_3   &  \lambda^3\cos\theta  &   \lambda^2\cos\theta  &
 \sin\theta }\ ,
\label{downmassex}
\end{eqnarray}
where $v_d=\langle H({\bf5}^*)\rangle$ and $\theta$ is the Higgs mixing 
angle. Our convention for the Dirac mass matrix $M_{ij}$ is that the 
fermion mass term is given by ${\cal L}_m = 
-(\overline{\psi_R}_iM_{ij}{\psi_L}_j +
\hbox{h.c.})$.\footnote{The
transpose T attached to $M_d$ in Eq.~(\ref{downmassex})
is due to this convention. Since
$\psi\cdot\chi^c\equiv\psi^\alpha\chi^c_\alpha=\chi^c\cdot\psi$ in
two-component notation
is equivalent to 
 $\overline{\chi_R}\psi_L$ in the usual four-component notation,
our Dirac mass term $\overline{\psi_R}_iM_{ij}{\psi_L}_j$ equals
$\psi^c_i\cdot M_{ij}\psi_j=\psi_i\cdot M^{\rm T}_{ij}\psi^c_j$, and
hence it corresponds to
$\int d^2\theta(\phi^c_iM_{ij}\phi_j)=\int d^2\theta(\phi_iM^{\rm
T}_{ij}\phi^c_j)$
written in terms of chiral superfields $\phi$ and $\phi^c$. This explains the
transpose in Eq.~(\ref{downmassex}).}
Note that the mixing terms retained in Eq.~(\ref{eq:eigen5star}),
although small, also contribute the same order amounts as indicated in 
 Eq.~(\ref{downmassex}) to  the $j1$ and $j2$ ($j=1,2,3$) matrix elements.
So they should be taken into account properly if we wish to calculate
not only the orders of magnitude but also the coefficients of the matrix
elements correctly. If we assume  $\sin\theta\sim\lambda^2$, then
\begin{eqnarray}
 M^{\rm T}_d=\lambda^2yv_d\pmatrix{
             \lambda^4  &  \lambda^3  &  \lambda^3  \cr
             \lambda^3  &  \lambda^2  &  \lambda^2  \cr
             \lambda   &   1    &   1    \cr}\ .
\label{downmasseex}
\end{eqnarray}
Thus we have $\lambda^2yv_d=m_b$. This factor $\lambda^2$ comes from the Higgs
mixing
$\sin\theta$ in the present E-twisted structure and provides a natural
explanation for the reason that the bottom mass is very small compared
with the top quark mass, in the small $\tan\beta\sim1$ scenario.

This mass matrix $M_d$ is common to the down quark
and lepton sectors at the GUT scale in the present approximation, and
it keeps almost the same form (aside from the over all normalization)
down to our low energy regime,  even after the
renormalization group evolution; so we have
$M_l \approx (m_\tau/m_b)M^{\rm T}_d$. (The reason that the transposition T
appears is because the left-handed and right-handed
components are contained in the $SU(5)$ ${\bf5^*}$
and ${\bf10}$ in opposite ways  for  the lepton and down-quark cases.)
It is important that the 32
and 33 elements have become of the same order. This gives a large mixing
angle in the charged lepton sector,
\begin{eqnarray}
M^\dagger_lM_l  = (m_\tau/m_b)^2 M_d^* M_d^{\rm T} =
m_\tau^2\times\pmatrix{
             \lambda^2  &  \lambda  & \lambda \cr
             \lambda   &   1   &  1  \cr
             \lambda   &   1   &  1  \cr }\ ,
\label{upmasssq}
\end{eqnarray}
and small mixing angles for down quark sector,
\begin{eqnarray}
  M^\dagger_dM_d =
m_b^2\times\pmatrix{
             \lambda^6   &  \lambda^5   & \lambda^3 \cr
             \lambda^5   &  \lambda^4   & \lambda^2 \cr
             \lambda^3   &  \lambda^2   &  1   \cr }\ .
\label{neumasssq}
\end{eqnarray}

The CKM and Maki-Nakagawa-Sakata (MNS) matrices\cite{ref:MNS} are given 
by
\begin{equation}
 U_{\rm CKM}=U_uU_d^\dagger\ , \quad \qquad  
 U_{\rm MNS}=U_lU_\nu^\dagger\ ,
\end{equation}
with $U_u,\ U_d,\ U_l$ and $U_\nu$  which make the matrices
$ U_u(M_u^\dagger M_u)U_u^\dagger, \ U_d(M_d^\dagger M_d)U_d^\dagger$,
$U_l(M_l^\dagger M_l)U_l^\dagger$
and $U_\nu^*M_\nu U_\nu^\dagger$ diagonal. The matrix $M_\nu$ is the
Majorana mass matrix of the light (almost) left-handed neutrinos,
which we discuss in the next section.

The remarkable fact is that the mass matrix in the down sector is not
symmetric even in such a large unification group as $E_6$ and that it
leads to quite different lepton and down quark mixing angles.
This results from the E-twisted structure of the third
generation. We have shown that the mass matrix form
(\ref{downmasseex}) surely reproduces the usual small down quark
mixings on the one hand and very large lepton 2-3 mixing on the
other. However, to show that the MNS matrix really gives the large
2-3 mixing, we must examine the the neutrino mass matrix $M_\nu$. 
We now turn to this task.

\section{Neutrino masses in E-twisted structure}

The Majorana mass matrix for the light neutrino is given by
\begin{equation}
  M_\nu= M_D^{\rm T}\,M_R^{-1}M_D\ ,
\end{equation}
where $M_D$ is the neutrino Dirac mass matrix and $M_R$ is
the Majorana mass matrix for the right-handed neutrinos.\cite{seesaw}

The neutrino Dirac masses come from the same Yukawa coupling 
Eq.~(\ref{upmass}) as the up quark masses:
\begin{eqnarray}
\Psi_{\bf 27}\Psi_{\bf 27}H_{\bf 27}
&&\ =
H({\bf 10},{\bf 5})\Big(
\Psi({\bf16},{\bf10})\Psi({\bf16},{\bf10}) \ \ \ \qquad \rightarrow\ \hbox{up
quark mass} \nn
&&\qquad \qquad \qquad\
{}+\Psi({\bf16},{\bf5}^*)\Psi({\bf16},{\bf1})\qquad \rightarrow\quad \nu\nu ^c
\nn
&&\qquad \qquad \qquad\
{}+\Psi({\bf10},{\bf5}^*)\Psi({\bf1})  \quad \qquad\ \ \rightarrow\quad N S
\quad \Big)\ .
\end{eqnarray}
Since our three left-handed neutrinos are
$\nu_1\in\Psi_1({\bf16},{\bf5}^*)$, $\nu_2\in\Psi_2({\bf16},{\bf5}^*)$ and
$N_3\in\Psi_3({\bf10},{\bf5}^*)$ in the present E-twisted
structure,\footnote{Precisely stated, there are $O(\lambda^3)$ and 
$O(\lambda^2)$
mixings of the component $\nu_3$ to the first and second generation
left-handed neutrinos, respectively, as indicated in
Eq.~(\ref{eq:eigen5star}). Here, however, we neglect this since the
mixing only affects the coefficients and does not change the
order of magnitude of the matrix elements for  the same reason as with 
 the down quark masses.}
the corresponding three right-handed neutrinos, which become their Dirac
mass partners, are
\begin{equation}
({\nu_1^c},\ {\nu_2^c},\ S_3)\ .
\end{equation}
Thus the Dirac mass matrix is found to take the form
\begin{equation}
(\nu_1^c \ \nu_2^c \ S_3)
M_{\rm D}
\pmatrix{\nu_1 \cr \nu_2 \cr N_3 \cr} \quad \hbox{with} \quad
M_{\rm D}= {\tilde m}_t
\pmatrix{
\lambda^6 & \lambda^5 &  0 \cr
\lambda^5 & \lambda^4 &  0 \cr
  0  &   0  &  1 \cr}\ , 
\label{eq:Diracmass}
\end{equation}
which is in parallel with the up quark mass matrix, 
except for the 3rd generation,  and exhibits a hierarchical structure.
The coefficient $\tilde m_t$ is $\sim m_t/3$,  where the factor 1/3 represents
the effect of renormalization group flow.

Next we study the right-handed neutrino Majorana masses,
which come from the higher dimensional interactions:
\begin{equation}
M_P^\m1\Psi_i({\bf 27})\Psi_j({\bf 27})
\bar\Phi^{\QQ_k}(\overline{\bf 27})\bar\Phi^{\QQ_l}(\overline{\bf 27})
(\Theta/M_P)^{\ff_i+\ff_j+\QQ_k+\QQ_l},
\end{equation}
where $\bar\Phi^{\QQ_k}(\overline{\bf 27})$ stands for
$\bar\Phi^{\QQ{=}{+}4}(\overline{\bf 27})$ and
$\bar H^{h{=}0}(\overline{\bf 27})$,
the partners of our Higgs fields $\Phi^{\QQ{=}{-}4}({\bf 27})$
and $H^{\QQ{=}0}({\bf 27})$.

Suppose that the $SO(10)$ singlet component of $\bar\Phi^{\QQ=4}$ and 
the $({\bf 16, 1})$ component of $\bar H^{\QQ{=}0}$ develop VEVs:
\begin{equation}
\left<\bar\Phi^{\QQ=4}(S)\right> = M_\Phi\ ,
\qquad
\left<\bar H^{\QQ=0}(\nu^c)\right> = M_H\ .
\end{equation}
All of the six right-handed neutrinos
$(\nu_i^c,\ S_i)$ $(i=1,2,3)$ acquire superheavy masses from this
interaction and mix with one another.
Now define the $2\times2$ matrix
\begin{equation}
\pmatrix{
 M_H^2       & M_HM_\Phi\lambda^4 \cr
 M_HM_\Phi\lambda^4 & M_\Phi^2\lambda^8  \cr}
\equiv
\pmatrix{
 M_1^{2} & M_1M_2 \cr
 M_1M_2  & M_2^2  \cr}\ .
\label{eq:2by2}
\end{equation}
Then, the $6\times6$ Majorana mass matrix for the
six right-handed neutrinos $(\nu_i^c,\ S_i)$ $(i=1,2,3)$ can be written
in the following tensor product form:
\begin{eqnarray}
M_\R&=&
M^\m1_\pl \times\bordermatrix{
     & \nu^c     & S   \cr
\nu^c &  M_1^{2} & M_1M_2 \cr
S    &  M_1M_2  & M_2^2 \cr}
\otimes
\bordermatrix{
   &  1   &   2  &   3  \cr
1  & \lambda^6 & \lambda^5 & \lambda^3 \cr
2  & \lambda^5 & \lambda^4 & \lambda^2 \cr
3  & \lambda^3 & \lambda^2 &   1  \cr} \nn
&=& {M_2^2\over M_\pl}\times
\bordermatrix{
     & \nu^c   &  S   \cr
\nu^c & \alpha^{2} &  \alpha \cr
S    &   \alpha  &  1   \cr}
\otimes
\bordermatrix{
   &  1   &   2  &   3  \cr
1  & \lambda^6 & \lambda^5 & \lambda^3 \cr
2  & \lambda^5 & \lambda^4 & \lambda^2 \cr
3  & \lambda^3 & \lambda^2 &   1  \cr}\ . 
\qquad
\left(\alpha\equiv{M_1\over M_2} \right)
\end{eqnarray}
The inverse of this matrix is given by
\begin{eqnarray}
M^\m1_\R&=&
{M_\pl\over M_2^2}\times
\bordermatrix{
     & \nu^c    &   S     \cr
\nu^c & \alpha^{-2} & \alpha^{-1} \cr
S    & \alpha^{-1} &   1     \cr}
\otimes
\bordermatrix{
   &  1   &   2  &   3  \cr
1  & \lambda^{-6} & \lambda^{-5} & \lambda^{-3} \cr
2  & \lambda^{-5} & \lambda^{-4} & \lambda^{-2} \cr
3  & \lambda^{-3} & \lambda^{-2} &   1    \cr}\ ,
\end{eqnarray}
so that the $3\times3$ submatrix for the three right-handed neutrinos
$({\nu_1^c},\ {\nu_2^c},\ S_3)$, which are Dirac mass partners of
our left-handed neutrinos $({\nu_1},\ {\nu_2},\ N_3)$, reads
\begin{equation}
M^\m1_{\R\,3\times3} =
{M_\pl\over M_2^2}\times
\bordermatrix{
        & \nu^c_1         & \nu^c_2         &  S_3           \cr
\nu^c_1  & \alpha^{-2}\lambda^{-6} & \alpha^{-2}\lambda^{-5} &
\alpha^{-1}\lambda^{-3} \cr
\nu^c_2  & \alpha^{-2}\lambda^{-5} & \alpha^{-2}\lambda^{-4} &
\alpha^{-1}\lambda^{-2} \cr
  S_3   & \alpha^{-1}\lambda^{-3} & \alpha^{-1}\lambda^{-2} &        1
\cr}\ .
\end{equation}
{}From this and Eq.~(\ref{eq:Diracmass}),
we find the induced left-handed Majorana mass matrix $M_\nu$ to be 
\begin{equation}
M_\nu= M_{\rm D}^{\rm T}\,M^\m1_{\R\,3\times3}M_{\rm D}
=
{{\tilde m}_t^2M_\pl\over M_2^2}\times
\bordermatrix{
      & \nu_1       & \nu_2       &  N_3     \cr
\nu_1  & \alpha^{-2}\lambda^{6} & \alpha^{-2}\lambda^{5} &
\alpha^{-1}\lambda^{3} \cr
\nu_2  & \alpha^{-2}\lambda^{5} & \alpha^{-2}\lambda^{4} &
\alpha^{-1}\lambda^{2} \cr
N_3   & \alpha^{-1}\lambda^{3} & \alpha^{-1}\lambda^{2} &        1
\cr}\ .
\end{equation}
If the parameter $\alpha=M_1/M_2$ is larger than $\lambda^2$, this 
left-handed neutrino mass matrix $M_\nu$ exhibits hierarchical 
structure, implying small mixing angles in the left-handed neutrino 
sector. 
Therefore the large 2-3 mixing angle in the MNS matrix suggested in the 
recent atmospheric neutrino experiment can be explained for a wide range
of parameters.

Finally, we add some numerology for the absolute values of light neutrino
masses. Suppose that
\begin{eqnarray}
&&M_2 = M_\Phi\lambda^4 \ \sim\  5\times10^{18}\,{\rm GeV}
\times(2\times10^{-3})
= 10^{16}\,{\rm GeV}\ , \nn
&&M_1 = M_H =\alpha M_2\ \sim\ \alpha\times10^{16}\,{\rm GeV}\ ,
\end{eqnarray}
which correspond to the following reasonable orders of VEVs:
\begin{eqnarray}
&&\left<\bar\Phi^{\QQ=4}(S)\right> = M_\Phi\ \ \sim\ 5\times10^{18}\,{\rm GeV}
\sim M_\pl\ ,
\nn
&&\left<\bar H^{\QQ=0}(\nu^c)\right> = M_H \ \sim\ \alpha\times10^{16}
\,{\rm GeV}\ .
\end{eqnarray}
If we assume  $\alpha\sim\lambda$, larger than $\lambda^2$, then,
with $M_\pl=10^{19}\,{\rm GeV}$ and ${\tilde m}_t\sim m_t/3\sim60\,{\rm GeV}$,
this yields the left-handed Majorana mass matrix $M_\nu$,
\begin{equation}
M_\nu=
\left({{\tilde m}_t^2M_\pl\over M_2^2}\ \sim\ 4\cdot10^{-2}{\rm
eV}\right)\times
\bordermatrix{
      & \nu_1   & \nu_2   &  N_3     \cr
\nu_1  & \lambda^{4} & \lambda^{3} & \lambda^{2} \cr
\nu_2  & \lambda^{3} & \lambda^{2} & \lambda^{1} \cr
N_3   & \lambda^{2} & \lambda^{1} &   1    \cr}\ ,
\end{equation}
with the eigenvalues
\begin{eqnarray}
m_{\nu_3}&\sim&4\times10^{-2}\,{\rm eV}\ , \nn
m_{\nu_2}&\sim&\lambda^2m_{\nu_3}\sim2\times10^{-3}\,{\rm eV}\ , \nn
m_{\nu_1}&\sim&\lambda^4m_{\nu_3}\sim1\times10^{-4}\,{\rm eV}\ .
\end{eqnarray}
These mass eigenvalues are consistent with the experimental data.

In the above, the $\alpha\sim\lambda$ was put by hand. There is another
option which gives this factor more naturally. This  is to adopt
Higgs fields $\Phi^{\QQ}$ and $\bar\Phi^{-\QQ}$ that carry $\QQ=-5$ 
instead of  $\QQ=-4$ as in the above scenario. 
This change causes essentially no change to
the mass matrix structure for the down quarks and charged leptons,
but it leads to some difference in the neutrino sector. In this case,
we must interchange the roles of $\bar\Phi^{-\QQ}$ and $\bar H^{\QQ=0}$;
namely, the $SO(10)$ $({\bf 16, 1})$ component of $\bar\Phi^{\QQ=5}$ and
the singlet component of $\bar H^{\QQ{=}0}$ develop VEVs:
\begin{eqnarray}
&&\left<\bar\Phi^{\QQ=5}(\nu^c)\right> = M_\Phi\ \
\sim\ 5\times10^{18}\,{\rm GeV} \sim M_\pl\ ,
\nn
&&\left<\bar H^{\QQ=0}(S)\right> = M_H \ \sim\ 10^{16}\,{\rm GeV}\ .
\end{eqnarray}
Then, the above $2\times2$ matrix in Eq.~(\ref{eq:2by2}) is replaced by
\begin{equation}
\bordermatrix{
     & \nu^c           & S           \cr
\nu^c &  M_\Phi^2\lambda^{10} & M_HM_\Phi\lambda^5 \cr
S    &  M_HM_\Phi\lambda^5   &  M_H^2      \cr}
\end{equation}
and leads to
\begin{eqnarray}
&& M_2 = M_H \ \sim\   10^{16}\,{\rm GeV}\ ,
 \nn
&& M_1 = M_\Phi\lambda^5 \ \sim\ 5\times10^{18}\,{\rm
GeV}\times(2\times10^{-3})\times\lambda
= \lambda\times10^{16}\,{\rm GeV}\ .
\end{eqnarray}
These values of $M_2$ and $M_1$ are the same as those found before.
Thus this also yields the same
left-handed Majorana mass matrix $M_\nu$ as in the previous case.
Note that $\alpha=M_1/M_2\sim\lambda$ was supplied from  one of the fifth
power $\lambda^5$, 
which came from the $U(1)_X$ charge assignment $\QQ=5$ to $\bar\Phi$.
One problem in this scenario which seems unnatural is, however,
that the $SO(10)$ $({\bf 16, 1})$ VEV of $\bar\Phi^{\QQ=5}$ must be 
larger than the $SO(10)$ singlet VEV of $\bar H^{\QQ{=}0}$,  
quite oppositely to the VEVs of their partner Higgs fields $H^{\QQ{=}0}$ 
and $\Phi^{\QQ=-5}$.

\section{Discussion and further problems}

We have presented a supersymmetric $E_6$ GUT model with anomalous
$U(1)_X$, which incorporates a novel mechanism
for yielding nonparallel mass structures between up-quark
and down-quark/charged-lepton sectors.

The hierarchical mass structure is basically explained by the
Frogatt-Nielsen mechanism using the $U(1)_X$ quantum numbers. However,
in large unification models with gauge groups larger than $SO(10)$, all 
 members of each generation fall into a single multiplet and hence
must carry a common value of the $U(1)_X$ charge. Therefore a
straightforward application of the Frogatt-Nielsen mechanism would yield
a {\it common} hierarchical mass matrix for all  up-, down- and
lepton sectors, in contradiction to  observations.
It seems that because of this `difficulty'  many authors who try to explain
the mass hierarchy with the Frogatt-Nielsen mechanism adopt smaller
unification gauge groups, or even discard the grand unification framework.

However,  we have proposed a novel mechanism which
can overcome this difficulty in an $E_6$ GUT framework.
We pointed out that there is a freedom of the $SU(2)_E$ inner automorphism
in $E_6$ in embedding $SO(10)$ in such a way that
$SU_{\rm GG}(5)\subset SO(10)\subset E_6$. As a reflection of this, the
fundamental
representation {\bf27} contains two $SU(5)$ ${\bf5}^*$ (and ${\bf1}$)
components, giving a spinor representation of $SU(2)_E$, and an
arbitrary linear combination of these two together with {\bf10} and a
{\bf1} can form a {\bf16} multiplet of an $SU(2)_E$-rotated $SO(10)$.
Our low energy fermions plus a right-handed neutrino just give an
$SO(10)$ {\bf16} at each generation. However, the $SO(10)$ groups 
chosen by three generations need not be the same, 
 but may be $SU(2)_E$ rotated from
one another. Moreover,  yet another $SU(2)_E$ rotation can also occur  in
the Higgs {\bf27}. Based on this observation, we examined the simplest
and typical version in which the $SO(10)$ groups chosen by the first and
second families coincide,  but the $SO(10)$ group of the third family 
is rotated by an angle $\pi$ from it.

It is remarkable that this simplest E-twisted scenario can reproduce all
the characteristic features of the fermion masses, not only the Dirac
masses but also the Majorana masses, by introducing only two Higgs
fields $H({\bf27})$ and $\Phi^{\QQ=-4}({\bf27})$, paired with $\bar
H(\overline{\bf27})$ and $\bar \Phi^{\QQ=4}(\overline{\bf27})$. The small
$SU(2)_E$ rotation with $\sin\theta\sim\lambda^2$ in the Higgs $H$ sector
explains
why the bottom quark is $\lambda^2$ times lighter than the top quark and why
the mixing between the second and third generation neutrinos is very
large. The large mass scales  $M$ and $M'$ for the superheavy fermions are
supplied by the two Higgs fields $H$ and $\Phi^{\QQ=-4}$, and their
partners $\bar H$ and $\bar \Phi^{\QQ=4}$ at the same time give the right
handed neutrino masses.

Here we did not use any higher representations for Higgs fields.
Note, however, that ${\bf 27}$ of $E_6$ contains no $SU(5)$ nonsinglet
component that is a standard gauge group singlet. Therefore, to break $SU(5)$
symmetry, we need at least one Higgs field of higher representations,
as long  as we stick to the conventional GUT framework. Another possibility
is to use the symmetry breaking mechanism by Wilson lines, as suggested
by string theory.
We also did not discuss  the detailed differences between the down-quark
and charged-lepton mass matrices. If one wishes to utilize a Georgi-Jarlskog
type mass matrix, one needs a  {\bf 45} representation of $SU(5)$,  which
is not included in ${\bf 27}$.

\section*{Acknowledgements}
The authors would like to thank Takeo Matsuoka, Naotoshi Okamura, Joe
Sato, Tsutomu Yanagida and Koichi Yoshioka for stimulating discussions.
They also thank the Summer Institute '98 held at Kyoto, where this
work was inspired by the stimulating seminars and discussions there.
One of the authors (M.\ B.) would like to thank the Aspen Center for
Physics for its hospitality and Pierre Ramond for valuable discussions.
M.~B.\ and T.~K.\ are supported in part by the Grants-in-Aid for
Scientific Research No.\ 09640375 and No.\ 10640261, respectively, from
the Ministry of Education, Science, Sports and Culture, Japan.

\appendix
\def\wbar#1{\overline{#1}}
\def\gam#1{\,{}^{#1}\kern-1pt\Gamma}
\def\sgm#1{\,{}^{#1}\!\sigma}
\def\barsgm#1{\,{}^{#1}\!\bar\sigma}
\def\Sgm#1{\,{}^{#1}\!\Sigma}
\def\CC#1{C_{(#1)}}
\def\dim#1{\,{}^{#1}\!}
\def\uni#1{{\bf 1}_{#1}}

\section{Representation {\bf27} and Maximal
Subgroups of $E_6$}

The adjoint representation {\bf78} of $E_6$ is decomposed under
its maximal subgroup $SO(10)\times U(1)_\VT$ as
${\bf 78}={\bf16}_{-3}+\wbar{\bf16}_{3}+{\bf45}_0+{\bf1}_4$
(The suffices denote the values of the $U(1)_\VT$ charge),
so that the $E_6$ generators are given by $SO(10)$ ${\bf16}$ Weyl-spinor
generators $E_\alpha$ $(\alpha=1,\cdots,16)$ and their complex conjugates
$\wbar E^\alpha
=(E_\alpha)^\dagger$ in addition to the 45 $SO(10)$ generators $T_{MN}$ and one
$U(1)$ generator $T$. The algebra is given by\cite{ref:IKK}
\begin{eqnarray}
[ \,T_{MN}, \,T_{KL}\, ] &=& -i\big( \delta_{NK}T_{ML}
+\delta_{ML}T_{NK} -\delta_{MK}T_{NL} -\delta_{NL}T_{MK} \big) \ ,
\nn
{}\bigl[ \,T_{MN}, \,\myvector{E_\alpha}{\wbar E^\alpha}\,\bigr]
&=& - \mymatrix{(\sigma_{MN})_\alpha^{\ \beta}}{0}{0}{(-\sigma
_{MN}^*)^\alpha_{\ \beta}}
\myvector{E_\beta}{\wbar E^\beta} \ , \nn
{}\bigl[ \,T, \,\myvector{E_\alpha}{\wbar E^\alpha}\,\bigr]
&=& {\sqrt3\over2}
\myvector{E_\alpha}{-\wbar E^\alpha} \ , \nn
{}[ \,E_\alpha, \,\wbar E^\beta\, ] &=& -{1\over2}(\sigma_{MN})_\alpha
^{\ \beta}T_{MN}
+ {\sqrt3\over2}\delta_\alpha^\beta T \ .
\label{eq:ALGEBRA}
\end{eqnarray}
(The $U(1)_\VT$ charge $\VT$ is related to $T$ by $\VT=2\sqrt3T$, and $T$
here is normalized in the same manner as  the other charges: 
${\rm Tr}(T^2)={\rm Tr}(T^2_{MN})={\rm Tr}(\wbar E^\alpha E_\alpha)$, 
where  $M,\ N$ and $\alpha$ are not summed over.)
The simplest representation {\bf 27} of $E_6$ is decomposed into ${\bf
1}_4 + {\bf 16}_{1} +{\bf 10}_{-2}$ under $SO(10)\times U(1)_\VT$, and
so it can
be denoted $\Psi_A\equiv(\psi_0, \psi_\alpha, \psi_M)$, where $\alpha$ and $M$
are  $SO(10)$ (Weyl-)spinor and vector indices, respectively. 
The $E_6$ generators act
on this representation as\cite{ref:IKK}
\begin{eqnarray}
&&\big(\theta T+{1\over2}\theta_{KL}T_{KL}+\bar\epsilon ^\gamma
E_\gamma+\wbar E^\gamma\epsilon _\gamma\big)
\pmatrix{\psi_0 \cr \psi_\alpha\cr \psi_M \cr} \nn
&&= \sanmatrix{{2\over\sqrt3}\theta}{\bar\epsilon ^\beta}{0}
{\epsilon_\alpha}
{{1\over2}\theta_{KL}(\sigma_{KL})_\alpha^{\
\,\beta}+{1\over2\sqrt3}\theta\delta_\alpha^{\ \beta}}
{-{1\over\sqrt2}(\sigma_NC\bar\epsilon^{\rm T})_\alpha}
{0}{-{1\over\sqrt2}(\epsilon^{\rm T}C\sigma^\dagger_M)^\beta}
{-i\theta_{MN}-{1\over\sqrt3}\theta\delta_{MN}}
\pmatrix{\psi_0 \cr \cr \psi_\beta\cr \cr \psi_N\cr}\ ,
\hspace{3em}
\label{eq:TWOSEVENREPR}
\end{eqnarray}
where $\sigma_{MN}\equiv(\sigma_M\sigma^\dagger_N-\sigma_N\sigma^\dagger_M)/4i$
is a $16\times16$
spinor representation matrix of the $SO(10)$ generators $T_{MN}$. Here 
$\sigma_M$ and $C$ are the $16\times16$ matrices with which the
$SO(10)$ $\gamma$-matrices $\Gamma_M$ and the charge conjugation matrix
$\CC{10}$,
satisfying $\Gamma_M\Gamma_N+\Gamma_N\Gamma_M=2\delta_{MN}$ and
$\CC{10}^{-1}\Gamma_M\CC{10}=\Gamma_M^{\rm T}$ and $\CC{10}=\CC{10}^{\rm T}$,
are given in the form
\begin{equation}
\Gamma_M=\pmatrix{0 & (\sigma_M)_{\alpha\beta} \cr
(\sigma_M^\dagger)^{\alpha\beta} & 0 \cr}, \qquad
\CC{10}=\pmatrix{0 & C^{{\rm T}\ \beta}_{\ \,\alpha}\cr C^\alpha_{\ \,\beta}& 0
\cr}\ .
\end{equation}
Note that $C\sigma_M^\dagger$ and $\sigma_MC$ are symmetric matrices.

An $E_6$ invariant can be constructed using three {\bf27}
representations  and is given
by  Eq.~(\ref{yukawacomp}),
$\Gamma^{ABC}\Psi_{1A}\Psi_{2B}\Psi_{3C}$,
where $\Gamma^{ABC}$ is a totally symmetric tensor:\cite{KugoSato}
\begin{equation}
\Gamma^{ABC} :\ \hbox{totally symmetric in $(A,\,B,\,C)$},
\quad \cases{ \Gamma^{0MN} = \delta_{MN}\ ,  \cr
\dsp \Gamma^{M\alpha\beta}=
{1\over\sqrt2}(C\sigma_M^\dagger)^{\alpha\beta}\ , \cr
\hbox{otherwise} \ \  0 \ \ .}
\end{equation}
By using a concrete representation for $\sigma_M$ and
 $C$ given in Ref.~\citen{KugoSato} 
in which the $SO(10)$ spinor {\bf16}, $\psi_\alpha$, is represented in 
 the form
\begin{equation}
\psi_\alpha({\bf16})
= \pmatrix{ \pmatrix{u_j \cr d_j \cr} \cr
  \rule[-2ex]{0pt}{5.5ex} \pmatrix{\nu\cr e \cr} \cr
 \rule[-2ex]{0pt}{5.8ex}\pmatrix{d^{c\,j} \cr -u^{c\,j}} \cr
  \pmatrix{e^c \cr -\nu^c \cr} \cr}\ ,
\end{equation}
the explicit form of the $SO(10)$ invariant ${\bf 16\times10\times16}$,
$\psi_{1\alpha}^{\rm T}(C\,H_M\cdot\sigma_M)^{\alpha\beta}\psi_{2\beta}$,
which appears in $\Gamma^{ABC}\Psi_{1A}\Psi_{2B}H_{C}$ (here 
$\Psi_{3C}({\bf27})$
is replaced by $H_C({\bf27})$  for clarity) is given as
\begin{eqnarray}
&&
\psi_1^{\rm T}({\bf16})C\big(H_M({\bf10})\cdot
{\sigma_M\over\sqrt2}\big)\psi_2({\bf16}) \nn
&&\quad =
\Bigl[\,
 (u^{c\,i}_1H^{\rm T}\epsilon^{\rm T}
  +d^{c\,i}_1H^{\prime\dagger})\pmatrix{u_i\cr d_i\cr}_2
 +(\nu^c_1H^{\rm T}\epsilon^{\rm T}
  +e^c_1H^{\prime\dagger})\pmatrix{\nu\cr e\cr}_2 \nn
&&\qquad\qquad {}+ H_i(e^c\ -\nu^c)_1\pmatrix{u^{c\,i}\cr d^{c\,i}\cr}_2
 -\bar H^{\prime\,i}(e\ -\nu)_1\pmatrix{u_i\cr d_i\cr}_2 \nn
&&\qquad\qquad \ {}+\varepsilon_{ijk}\bar H^{\prime\,i}u_1^{c\,j}d_2^{c\,k}
           -\varepsilon^{ijk} H_iu_{1\,j}d_{2\,k}\
\Bigr] \,\  + \, \ (1\leftrightarrow2)\ ,
\label{eq:161016}
\end{eqnarray}
where $i,\,j$ and $k$ denote $SU(3)$ color,  and
the components of $H_M({\bf10})$ are denoted by
\begin{eqnarray}
&&H({\bf5}) = \pmatrix{\cspan H_i \cr H \cr}\,,\quad
\cases{
\rule[-2ex]{0pt}{1ex}
H_i={i\over\sqrt2}(H_{M=i+3}+iH_{M=i})\ ,\quad (i=1,2,3) \cr
H=\pmatrix{H_x \cr H_y \cr}
= {1\over\sqrt2}\pmatrix{-iH_7-H_8 \cr H_0+iH_9 \cr}\,, \cr} \nn
&&H({\bf5}^*) = \pmatrix{\cspan\bar H^{\prime\,i} \cr \bar H'\cr}\,, \quad
\cases{
\rule[-2ex]{0pt}{1ex}
\bar H^{\prime\,i}=-{i\over\sqrt2}(H_{M=i+3}-iH_{M=i})\ ,\quad (i=1,2,3) \cr
\bar H'=\pmatrix{\bar H'_x \cr \bar H'_y \cr}
= {1\over\sqrt2}\pmatrix{iH_7-H_8 \cr H_0-iH_9 \cr}\,. \cr}
\end{eqnarray}
These $SU(5)$ {\bf5} and ${\bf5}^*$ components,
$H({\bf5})$ and $H({\bf5}^*)$,
correspond to the decomposition
${\bf10}={\bf5}_2+{\bf5}_{-2}^*$ given in Eq.~(\ref{eq:su510}) for
$\psi_M({\bf10})$. Then the $SO(10)$ invariant $H_{M}\psi_{M}$ is given
\begin{equation}
\sum_{M=1}^{10}H_{M}({\bf10})\psi_{M}({\bf10})=
\pmatrix{\ \bar H^{\prime\,i}&H^{\prime \dagger}\cr}
\pmatrix{\cspan D_i \cr \pmatrix{E^c \cr -N^c \cr} \cr} +
\pmatrix{\ D^{c\,i} & (\,E,\ -N\,) \cr}
\pmatrix{\cspan H_i \cr H \cr}\ .
\end{equation}
In connection with this $SU(5)$ decomposition, we note that
the above $SO(10)$ invariant ${\bf 16\times10\times16}$ given in
Eq.~(\ref{eq:161016}) is expressed as
\begin{eqnarray}
&&\psi_1^{\rm T}({\bf16})C\big(H_M({\bf10})\cdot
{\sigma_M\over\sqrt2}\big)\psi_2({\bf16})
={1\over2^2}\varepsilon^{ijklm}\psi_{1\,ij}({\bf10})
\psi_{2\,kl}({\bf10})H_m({\bf5}) \nn
&&\qquad \quad  +\left\{\left(
\psi_{1\,ij}({\bf10})\bar H'^i({\bf5}^*)\psi_2^j({\bf5}^*)
-\nu^c_1H_i({\bf5})\psi_2^i({\bf5}^*)\right) +
\ (1\leftrightarrow2) \right\}
\end{eqnarray}
in terms of the $SU(5)$ components 
 $\psi_{1\,ij}({\bf10})$, $\psi^i({\bf5}^*)$
and $\nu^c({\bf1})$ contained in $\psi_\alpha({\bf16})$:
\begin{equation}
\psi_{ij}({\bf10}) =
\pmatrix{\cspan \varepsilon_{ikj}u^{c\,k} & -u_i & -d_i \cr
                   u_j           &  0   & -e^c \cr
                   d_j           &  e^c &   0  \cr}\ ,\qquad
\psi^i({\bf5}^*) = \pmatrix{\cspan d^{c\,i} \cr e \cr -\nu\cr}\ .
\label{eq:su5comp}
\end{equation}

Under a maximal subgroup $SU(3)_L\times SU(3)_R\times SU(3)_c\subset E_6$,
$\Psi_A({\bf27})$ is decomposed into the following three irreducible
components
\begin{eqnarray}
&&\psi_{i_Lj_c}({\bf 3,1,3}) =
 \pmatrix{u_j\cr d_j\cr D_j\cr}\ , \qquad
\psi^{i_cj_R}({\bf 1,3^*,3^*})=
 (u^{c\,i} \  d^{c\,i} \ D^{c \,i})\ , \nn
&&\psi_{i_R}^{\ \ j_L}({\bf 3^*,3,1})
=\bordermatrix{      &  1^*_L & 2^*_L  & 3^*_L \cr
                1_R  &  N^c & E^c  & -e^c \cr
                2_R  &  -E  &  N   & \nu^c \cr
                3_R  &   e  & -\nu &  -S  \cr}\ ,
\label{eq:333decomp}
\end{eqnarray}
where the suffices $L,\,R,\,c$ are attached to the indices $i,\,j$ to
distinguish which of $SU(3)$s  the indices refer to, and the
component fields are those defined in Eqs.~(\ref{eq:su516})
-- (\ref{eq:su1}).
In terms of these $SU(3)_L\times SU(3)_R\times SU(3)_c$ component fields,
the $E_6$ invariant (\ref{yukawacomp}),
$\Gamma^{ABC}\Psi_{1A}\Psi_{2B}\Psi_{3C}$,
trilinear in {\bf27} can be written in the form:
\begin{eqnarray}
&&\hspace{-1em}\Gamma^{ABC}\Psi_{1A}\Psi_{2B}\Psi_{3C}=
-\bigl(\psi_1^{i_cj_R}
\psi_{2\,j_R}^{\ \ \ k_L}
\psi_{3\,k_Li_c}
+ (\hbox{$3!-1$ permutations of } (1,2,3))\bigr)
\nn
&& \qquad
{}-\varepsilon^{i_Lj_Lk_L}\varepsilon^{l_cm_cn_c}
\psi_{1\,i_Ll_c}\psi_{2\,j_Lm_c}\psi_{3\,k_Ln_c}
+\varepsilon_{i_Rj_Rk_R}\varepsilon_{l_cm_cn_c}
\psi_1^{l_ci_R}\psi_2^{m_cj_R}\psi_3^{n_ck_R}
\nn
&& \qquad \qquad
{}-\varepsilon^{i_Rj_Rk_R}\varepsilon_{l_Lm_Ln_L}
\psi_{1\,i_R}^{\ \ \ l_L}\psi_{2\,j_R}^{\ \ \ m_L}
\psi_{3\,k_R}^{\ \ \ n_L}\ .
\end{eqnarray}

Under another maximal subgroup $SU(6)\times SU(2)_E\subset E_6$ discussed in
Eq.~(\ref{eq:su6decomp}),
the $\Psi_A({\bf27})$ is decomposed into the following two irreducible
components: 
\begin{equation}
\psi_{ij}({\bf15},{\bf1}) =
\pmatrix{\cspan \varepsilon_{ikj}u^{c\,k} & -u_i & -d_i & -D_i \cr
                   u_j           &  0   & -e^c & -E^c \cr
                   d_j           &  e^c &   0  &  N^c \cr
                   D_j           &  E^c & -N^c &   0  \cr},\quad
\psi^i_{\ a}({\bf6^*},{\bf2}) =
\pmatrix{\cspan d^{c\,i} & -D^{c\,i}\cr e & -E \cr -\nu& N \cr -S &
                   \nu^c\cr}\ .
\label{eq:su6comp}
\end{equation}
In terms of these,
the same $E_6$ invariant (\ref{yukawacomp}) can be written in the form
\begin{eqnarray}
&&\hspace{-1em}\Gamma^{ABC}\Psi_{1A}\Psi_{2B}\Psi_{3C}=
-{1\over2^3}\varepsilon^{ijklmn}\psi_{1\,ij}\psi_{2\,kl}\psi_{3\,mn} \nn
&&\qquad {}+\bigl(\varepsilon^{ab}\psi_{1\,ij}\psi_{2\,a}^i\psi_{3\,b}^j
+ (\hbox{cyclic permutations of }(1,2,3))\bigr)\ .
\label{eq:su6su2inv}
\end{eqnarray}
It is clear from Eq.~(\ref{eq:su6comp}) that  $SU(2)_E$ is an
$SU(2)$ subgroup of $SU(3)_R$,  which acts on the second and third entries
of the fundamental representation {\bf3} of $SU(3)_R$.
Note also that $SU(5)_{\rm GG}$ and $SU(3)_L$ are regular subgroups of
this $SU(6)$ such that the first five entries
and the last three entries of the {\bf6} of $SU(6)$ are the fundamental
representations {\bf5} and {\bf3} of $SU(5)_{\rm GG}$ and $SU(3)_L$,
respectively.

Another interesting $SU(6)\times SU(2)$ subgroup, discussed by
Haba et al.,\cite{ref:matsuoka}  is
$SU(6)'\times SU(2)_R\subset E_6$, where
$SU(6)'\supset SU(4)_{{\rm Pati}\hbox{-}{\rm Salam}}\times SU(2)_L$ and
$SU(2)_R$ is
an $SU(2)$ subgroup of $SU(3)_R$ which acts on the first and second
entries of the fundamental representation {\bf3} of $SU(3)_R$.
Under this subgroup $SU(6)'\times SU(2)_R$,
 $\Psi_A({\bf27})$ is decomposed into the following two irreducible
components: 
\begin{equation}
\psi_{ij}({\bf15},{\bf1}) =
\pmatrix{\cspan -\varepsilon_{ikj}D^{c\,k} & -D_i & -u_i & -d_i \cr
                    D_j           &  0   &  \nu &   e  \cr
                    u_j           & -\nu &   0  &   S \cr
                    d_j           & -e   &  -S  &   0  \cr},\quad
\psi^i_{\ a}({\bf6^*},{\bf2}) =
\pmatrix{\cspan d^{c\,i} & -u^{c\,i}\cr -e^c & \nu^c \cr
                               N^c & -E \cr E^c & N \cr}\ .
\label{eq:su6pcomp}
\end{equation}
This subgroup $SU(6)'\times SU(2)_R$ can essentially be obtained from the above
$SU(6)\times SU(2)_E$ by an inner automorphism by an element $\exp(-i\pi
J_2^{R\prime})$,  where $J_2^{R\prime}$ is the second generator of  yet
another $SU(2)$ subgroup of $SU(3)_R$ which acts on the first and third
entries of the fundamental representation {\bf3} of $SU(3)_R$. Indeed,
under this rotation $\exp(-i\pi J_2^{R\prime})$, doublets {\bf2} of this
$SU(2)$ are changed as $(u,\,d) \rightarrow(-d,\,u)$,  so that the $SU(3)_R$
non-singlet components in the $SU(3)_L\times SU(3)_R\times SU(3)_c$
decomposition
Eq.~(\ref{eq:333decomp}) are transformed as
\begin{eqnarray}
(u^{c\,i} \  d^{c\,i} \ D^{c \,i})  \quad \qquad &\rightarrow& \quad \qquad
 (-D^{c\,i} \  d^{c\,i} \ u^{c \,i})\ , \nn
\bordermatrix{      &  1^*_L & 2^*_L  & 3^*_L \cr
                1_R  &  N^c & E^c  & -e^c \cr
                2_R  &  -E  &  N   & \nu^c \cr
                3_R  &   e  & -\nu &  -S  \cr}
 \quad  &\rightarrow& \quad 
\bordermatrix{      &  1^*_L & 2^*_L  & 3^*_L \cr
                1_R  &  -e  & \nu  &  S   \cr
                2_R  &  -E  &  N   & \nu^c \cr
                3_R  &  N^c & E^c  & -e^c \cr}\ .
\end{eqnarray}
The field content in Eq.~(\ref{eq:su6pcomp}) is actually obtained
from Eq.~(\ref{eq:su6comp}) with this replacement for the component fields
together with a relabeling of the 
last three indices of $SU(6)$, $(4,5,6) \rightarrow(6,4,5)$.
By this inner automorphism, it is clear that the $E_6$
invariant trilinear in {\bf27} is given by the same formula, 
Eq.~(\ref{eq:su6su2inv}),  also in this case.
(If one wishes to have an $SU(4)_{{\rm Pati}\hbox{-}{\rm Salam}}$ which has
$(u_i,\,\nu)$ and $(d^c_i,\,e^c)$ as its ${\bf4}$ and ${\bf4}^*$  instead
of $(u_i,\,-\nu)$ and $(d^c_i,\,-e^c)$ in Eq.~(\ref{eq:su6pcomp}),
one can easily obtain this by using another suitable inner automorphism by
an element of $SU(2)\subset SU(3)_L$, since $SU(2)$ spinors change sign under
$2\pi$ rotation.)

Finally, as is now clear, yet another 
$SU(6)\times SU(2)$ subgroup,\cite{ref:Kawamura}
$SU(6)''\times SU(2)'_R$, can be obtained from 
$SU(6)\times SU(2)_E$ by an
inner automorphism by a $\pi$ rotation $\exp(-i\pi J_2^R)$ of $SU(2)_R$.
Applying this transformation $\exp(-i\pi J_2^R)$ to Eq.~(\ref{eq:su6comp}),
we find that  $\Psi_A({\bf27})$ is decomposed under this
subgroup $SU(6)''\times SU(2)'_R$ into
\begin{equation}
\psi_{ij}({\bf15},{\bf1}) =
\pmatrix{\cspan -\varepsilon_{ikj}d^{c\,k} & -u_i & -d_i & -D_i \cr
                   u_j           &  0   & -\nu^c &  N  \cr
                   d_j           & \nu^c &   0  &   E  \cr
                   D_j           &  -N  &  -E  &   0  \cr},\quad
\psi^i_{\ a}({\bf6^*},{\bf2}) =
\pmatrix{\cspan u^{c\,i} & -D^{c\,i}\cr
e & N^c \cr -\nu& E^c \cr -S & -e^c \cr}\ .
\end{equation}
Clearly the $SU(5)$ subgroup of this $SU(6)''$, acting on  the first five
entries, is the flipped $SU(5)$.

\end{document}